\documentclass[allclo]{FBSart}
\usepackage{amsfonts}
\usepackage{amssymb}
\usepackage{graphicx,color}

\title{Coulomb interactions within Halo EFT}
\author{R. Higa\thanks{\textit{E-mail address:} 
higa@itkp.uni-bonn.de}}
\institute{Helmholtz-Institut f\"ur Strahlen- und Kernphysik (Theorie),
Universit\"at Bonn, Nu\ss allee 14-16, 53115 Bonn, Germany}

\runningauthor{R. Higa}
\runningtitle{Coulomb interactions within Halo Effective Field Theory}
\begin{document}

\maketitle
\begin{abstract}
I present preliminary results of effective field theory applied to 
nuclear cluster systems, where Coulomb interactions play 
a significant role. 
\end{abstract}

\section{Introduction}

Nuclear systems far from the so-called valley of stability brought 
a lot of excitement to the field in the last years. While traditional 
approaches like shell model and mean field techniques provide 
qualitative descriptions of stable nuclei, they are seriously challenged 
by isotopes closer to either proton or neutron drip lines. The latter 
tend to form clusters loosely bound among themselves. 
Many of these isotopes, in particular halo nuclei, exhibit large 
cross-section at low energies, which can be quite relevant to 
reaction rates in nuclear astrophysics. 

The weak binding of such cluster systems are usually 
well-separated from the next higher energy scale, for instance, the 
excitation energy of each cluster (nucleons and/or $\alpha$ particles). 
That turns out to be an attractive scenario for effective field theory 
(EFT) studies. The formalism takes into account only the relevant 
degrees of freedom at low momentum $k$ and, 
according to a defined set of rules (power counting) 
provides a controlled and systematic 
expansion of physical quantities in powers of $k/M_{hi}\sim M_{lo}/M_{hi}$, 
where $M_{lo}$ and $M_{hi}$ set the magnitude of low and high momenta 
scales \cite{eftrev1,eftrev2}. 

Halo EFT was introduced in \cite{BHvK} with application to neutron-alpha
scattering. Here I present applications of Halo EFT to alpha-alpha 
and proton-alpha systems, where Coulomb forces are important. 
These three basic interactions constitute the starting point for a 
description towards heavier nuclear systems. 

\section{$\alpha\alpha$ scattering}

At low energies ($E_{LAB}\lesssim 3$ MeV) $\alpha\alpha$ scattering 
is dominated by $S$-wave and characterized by the existence of a resonance 
at $E_R=184$ KeV and width $\Gamma_R=11$ eV (the ${}^8$Be ground 
state\footnote{This resonance is quite relevant to the triple-$\alpha$ 
process, leading to the synthesis of ${}^{12}$C in massive stars.}). 
Analyses of scattering data 
using effective range theory reveal an incredibly large scattering 
length, $a_0\sim 10^3$ fm \cite{ere}, thus implying that our power counting 
needs more fine-tuning than naively expected. In~\cite{HHvK} we developed a 
power counting for the $\alpha\alpha$, wich results in a very large 
scattering length, 
$a_0\sim M_{hi}/M_{lo}^2$, and a non-perturbative (but still of natural 
size) effective range $r_0\sim 1/M_{hi}$. 
Coulomb interactions were dealt non-perturbatively 
along the lines of \cite{clbeft} and the inverse of the amplitude 
becomes proportional to 
\begin{equation}
-1/a_0+r_0\,k^2/2-2H(\eta)/a_B\quad +\quad \mbox{subleading terms}\,,
\label{eq:aa1}
\end{equation}
where $a_B=2/(m_{\alpha}Z_{\alpha}^2\alpha_{em})\approx 137/(2m_{\alpha})$ 
is the $\alpha\alpha$ ``Bohr radius", $\eta=(a_Bk)^{-1}$ and 
$H(x)=\psi(ix)+(2ix)^{-1}-\ln(ix)$. 

Interesting in this power counting is that, when Coulomb interactions 
are turned off, the third term of Eq.~(\ref{eq:aa1}) becomes the usual 
unitarity term $-ik$, while the first two become subleading corrections. 
Therefore, at leading order the ${}^8$Be system shows conformal invariance, 
and the corresponding 3-body system ${}^{12}$C acquires an exact Efimov 
spectrum \cite{eftrev2}. This is a possible realization of the 
unitarity limit. When Coulomb is restored, the $1/r$ potential breaks 
scale invariance and the three terms of Eq.~(\ref{eq:aa1}) are of 
comparable size. However, the fact that the ${}^8$Be ground state stays 
close to threshold can be seen as a reminiscence from this broken 
unitary regime. 

\begin{figure}[ht]
\begin{center}
\includegraphics[height=4.5cm,width=7.0cm]{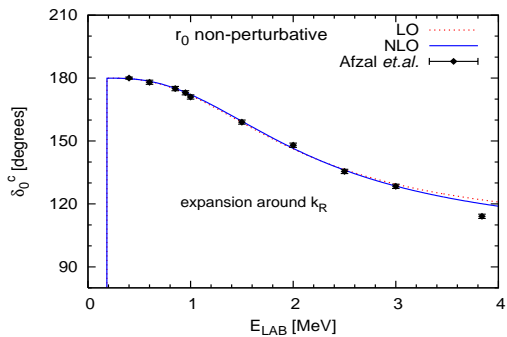}
\end{center}
\vspace*{-0pt}
\caption{$S$-wave phase shift for $\alpha\alpha$ scattering as a 
function of the laboratory energy $E_{LAB}$.}
\label{fig:aascatt}
\end{figure}

We fit our EFT expressions to the available $\alpha\alpha$ scattering 
data (Fig.~\ref{fig:aascatt}) and find agreement in the effective 
range parameters \cite{ere} except for $a_0$, whose inverse is very 
sensitive to 
big cancellations that occur between strong and electromagnetic 
contributions \cite{HHvK}. However, the order of magnitude is the same, 
which indicates a lot of fine-tuning in the $\alpha\alpha$ system that 
remains to be understood. 

\section{$p\alpha$ scattering}

\begin{figure}[ht]
\begin{center}
\includegraphics[height=4.5cm,width=7.0cm]{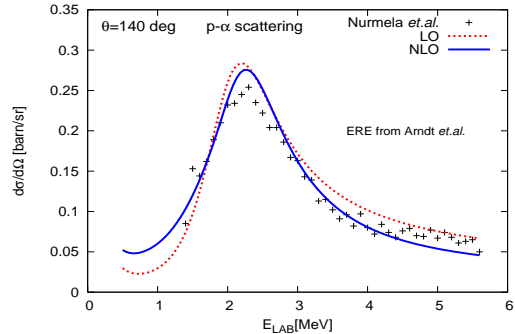}
\end{center}
\vspace*{-0pt}
\caption{Differential cross-section for $p\alpha$ scattering as a 
function of the laboratory energy $E_{LAB}$, at fixed angle 
$\theta=140^{\circ}$.}
\label{fig:pascatt}
\end{figure}

In $p\alpha$ scattering one is interested in low-energies 
$E_{LAB}\lesssim 4$ MeV \cite{BRvK}. Phase shift analysis from 
\cite{padat1} indicates that $S_{1/2}$, $P_{1/2}$, and $P_{3/2}$ are 
the dominant waves in this region, the latter showing a resonance 
around $E_{LAB}\sim 2.3$ MeV. We extended the formalism of \cite{clbeft} 
to include $P$-waves, and adopted the same power counting from 
Ref.~\cite{BHvK}, where the $P_{1/2}$ wave doesn't contribute up to NLO. 
Comparison with differential cross-section data from Ref.~\cite{padat2}, 
using the effective range parameters from \cite{padat1} as input, shows 
convergence and good agreement (Fig.~\ref{fig:pascatt}). 

\begin{acknowledge}
I would like to thank Hans-Werner Hammer, Bira van Kolck, and Carlos 
Bertulani for stimulating collaboration, and the organizers of the 
EFB20 for the excellent conference. This work was partially support 
by the BMBF under contract number 06BN411, and by DOE Contract
No.DE-AC05-06OR23177, under which SURA operates the Thomas Jefferson 
National Accelerator Facility. 
\end{acknowledge}


\begin{thebibliography}{9}

\bibitem{eftrev1} P.F.~Bedaque and U.~van Kolck,
Ann.\ Rev.\ Nucl.\ Part.\ Sci.\  {\bf 52}, 339 (2002).

\bibitem{eftrev2} E.~Braaten and H.-W.~Hammer,
Phys.\ Rept.\  {\bf 428}, 259 (2006). 

\bibitem{HHvK} R. Higa, H.-W. Hammer, and U. van Kolck, in preparation. 

\bibitem{clbeft} X.\ Kong and F.\ Ravndal, 
Phys. Lett. {\bf B450} 320 (1999); 
Nucl.\ Phys. {\bf A665}, 137 (2000); 
B. R. Holstein, 
Phys. Rev. D {\bf 60}, 114030 (1999). 

\bibitem{ere} J.L. Russell, Jr. {\em et. al.},
Phys. Rev. {\bf 104}, 135 (1956);
G. Rasche, Nucl. Phys. {\bf A94}, 301 (1967); 
S.A. Afzal {\em et. al.}, Rev. Mod. Phys. {\bf 41}, 247 (1969).

\bibitem{BHvK} C. A. Bertulani, H.-W. Hammer, and U. van Kolck, 
Nucl. Phys. {\bf A712}, 37 (2002);
P. F. Bedaque, H.-W. Hammer, and U. van Kolck, 
Phys. Lett. {\bf B569}, 159 (2003). 

\bibitem{BRvK}
C.A. Bertulani, R. Higa, and U. van Kolck, in progress. 

\bibitem{padat1} R.A.\ Arndt, L.D.\ Roper, and R.L.\ Shotwell,
Phys.\ Rev.\ C {\bf 3}, 2100 (1971).

\bibitem{padat2} A. Nurmela, E. Rauhala, and J. R\"ais\"anen, 
J. Appl. Phys. {\bf 82}, 1983 (1997).



\end{thebibliography}
\end{document}